# Personalized Visual Support After Automatic Speech Recognition

*Beatriz Calero[1], Duarte Neves[1,2], Carlos Bruno[3], Luis Ricardo Santos[3,4], Mario Egea[1,2]*

[1] Faculty of Engineering, University of Porto
[2] Faculty of Psychology and Education Sciences, University of Porto
[3] Faculty of Psychology and Education Sciences, University of Coimbra
[4] Institute of Intelligent Systems and Robotics, CNRS, Sorbonne Université
lrsantos@fpce.uc.pt, megea@fe.up.pt

## Abstract

Many spoken-language interfaces stop at producing a transcript. This project examines how a second layer of optional text and pictorial support can be adapted to individual needs. We specify four presentation levels and a behavior-based selection policy driven by replay requests, response latency and recent comprehension. After expert review of the mappings, we will compare fixed and adaptive presentation using counterbalanced spoken instructions. The study is designed to measure comprehension, usability and which modality listeners rely on, not to claim that transcript embellishment uniformly benefits listeners.

**Index Terms**: multimodal communication, adaptive interfaces, speech-to-text, pictograms

## 1. Introduction

A transcript is a useful accessibility feature, but it carries almost the same informational organization as the original utterance. The question considered here is whether a listener should always receive the same presentation or be offered an amount of structure that changes with task performance.

Speech recognition now supports practical transcription in many everyday situations, but a transcript does not by itself guarantee comprehension. Large-scale weak supervision has improved robust speech recognition across acoustic conditions [1]. For listeners dealing with rapid instructions, unfamiliar vocabulary or competing demands, the next problem is deciding which information to present and in what form.

The evidence on added text is mixed, so this decision cannot be settled once for everyone. A review of more than one hundred studies of video captioning reports gains in comprehension, attention and memory, particularly for viewers working in a non-native language, although these studies concern video rather than spoken task instructions [2]. Multimedia-learning research cautions against presenting identical printed and spoken streams, because processing capacity in each channel is limited [3]. The expertise reversal effect makes a related point: guidance that helps less knowledgeable learners can lose its value, or become counterproductive, as knowledge increases [4]. Taken together, these findings argue for support that depends on the listener's current need rather than staying constant.

Reading support studies suggest that sentence segmentation, pictograms and labels need not help every learner equally [5]. Automated text-to-pictogram mapping has been assessed in several languages and can supply candidate visual supports, though its reliability depends on lexical coverage and context [6]. Pictogram norming points the same way: in a study of 1,525 pictograms from the ARASAAC collection, meanings were on average hard to guess unaided yet were rated as closely related to their intended referents, with nouns scoring highest on both measures [7]. This supports pairing pictograms with text, and suggests that action pictograms need particularly careful review. The proposed project concerns the selection of these resources after speech recognition, rather than the recognition algorithm itself.

Differences in language-related processing should not be collapsed into a single severity scale. A variety of impairments were shown across different conditions [8], [9], [10], [11], [12]. Multiple studies made attempts to decode neural markers during language processing in such conditions [13], [14], [15], [16], [17], [18], as well as develop interfaces [19], [20] for patients to communicate [21] and interact with text [22], illustrating that access modality and understanding are distinct problems. Our task thus concerns comprehension of heard instructions, not decoding intended speech.

We address three questions. First, does behavior-adaptive presentation improve comprehension of action order relative to a fixed format? Second, does it do so without offering support that listeners then reject? Third, which source (audio, text or images) do listeners actually use to resolve each task?

## 2. System Concept

A spoken instruction will be transcribed and aligned with sentence or clause boundaries. The listener can receive the original transcript, a segmented transcript, a segmented transcript with selected pictograms, or a display emphasizing ordered action steps (Table 1). The acoustic recording will remain available for replay. Figure 1 shows how these parts connect and how observed behavior feeds back into the choice of level.

**Table 1.** *The four presentation levels.*

| Level | Display | Added over previous level | Reason for a separate level |
|---|---|---|---|
| 1 | Original transcript | Text aligned with the recording | Baseline; text may help or duplicate the audio |
| 2 | Segmented transcript | Clause or sentence breaks | Isolates segmentation from semantic support |
| 3 | Segmented transcript with pictograms | Reviewed icons for actions, objects and locations | Visual cue kept beside the label, since icons alone are often hard to guess |
| 4 | Ordered action steps | Segments laid out as a numbered sequence in the original wording | Targets the order errors counted in the primary endpoint |

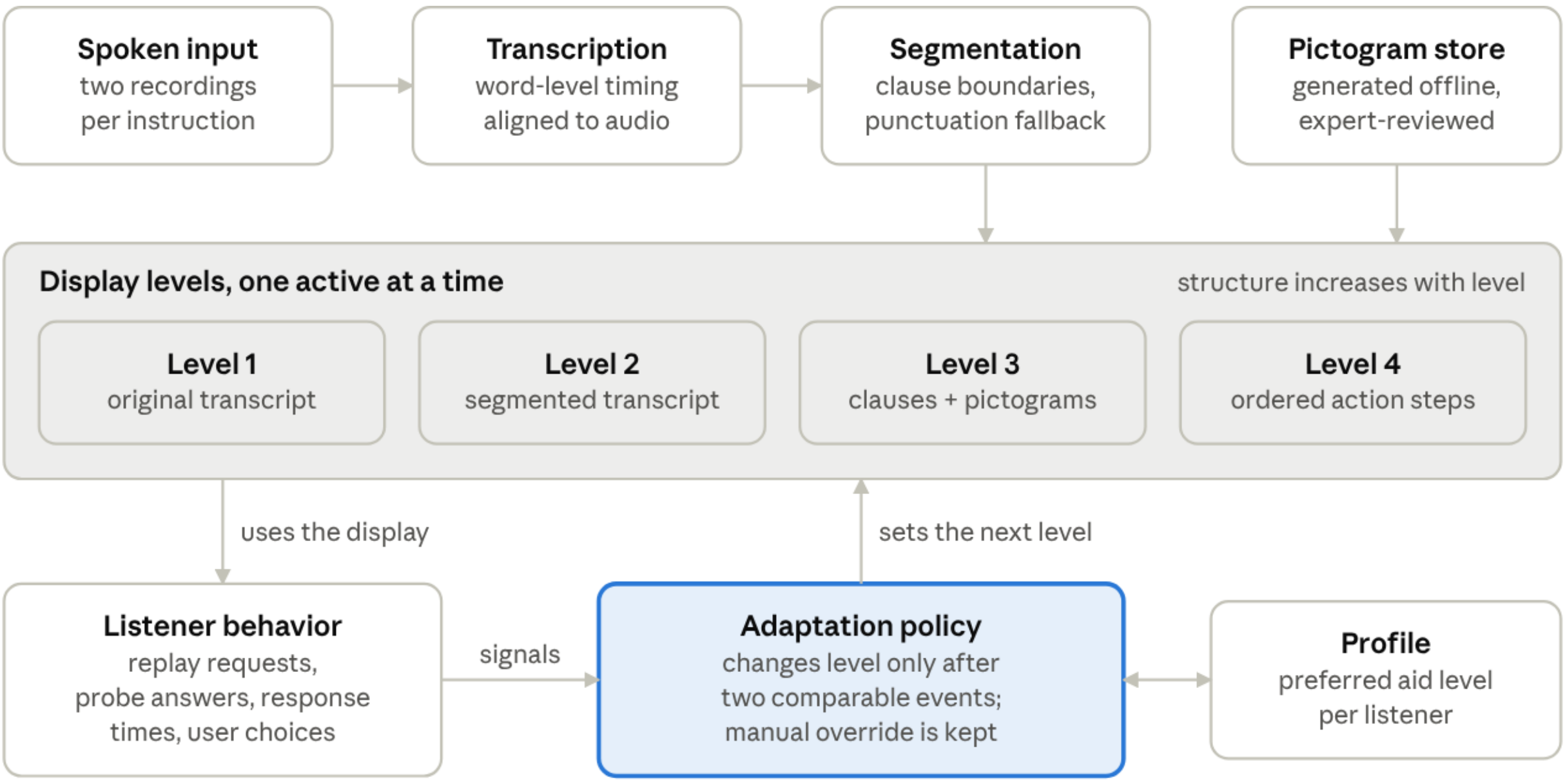

**Figure 1.** *System flow: prepared text and pictograms feed one active display level, and observed listener behavior updates that level through the adaptation policy.*

Selection will initially depend on observable behavior: replay requests, answers to a small comprehension probe, response times, and user choices. Each trial yields a difficulty event (a replay request, an incorrect probe answer, or a response slower than a threshold calibrated on the training set), an unaided success, or neither. A simple rule will offer a richer display following repeated difficulty and reduce it when the person succeeds without assistance. The reasoning follows ability-based design, where the system adjusts to demonstrated capabilities rather than requiring a fixed user type [23]; the rule is deliberately symmetric, because support that helps at one level of knowledge can become redundant at another [4], [5].

Because comprehension depends on integration across clauses, visually marking every concept may be counterproductive. The construction-integration account of discourse comprehension provides one reason to examine segmentation and semantic support separately [24]. Segmentation has a practical precedent in subtitling: in an eye-tracking study of live respoken subtitles, chunked text reduced the time participants spent reading [25]. Whether it also improves understanding is what Level 2 is meant to test. Icons will be limited to actions, objects and locations after human review for polysemy, a restriction consistent with pictogram norms in which word concreteness and imageability were the factors most strongly correlated with how readily pictograms were guessed and matched to their meanings [7].

A lightweight personalized profile will keep track of preferred aid levels. To prevent overreaction to a single slow response, the policy will change format only after two comparable events and will preserve a manual override. In the terms of Plass and Pawar's taxonomy of adaptivity [26], the policy thus combines adjustment during a task with a profile that accumulates over longer use. Such conservative adaptation is attractive for an initial prototype, where reliable estimation of individual preferences is more important than algorithmic complexity.

The transcription stage will preserve word-level timing so that text segments remain associated with the original recording. Whisper-style recognizers [1] do not provide reliable word-level timestamps out of the box, but voice-activity detection followed by forced phoneme alignment can supply them [27]. We will develop a conservative sentence and clause segmenter and evaluate it manually on the instructional materials, reporting how often it is uncertain. When segmentation is uncertain, the display will fall back to punctuation-defined units rather than invent a syntactic analysis.

Pictograms will be generated offline, checked by language specialists and stored with their source phrase and sense. The interface therefore tests support selection rather than conflating that test with stochastic changes in image selection.

Instructions will be sampled from everyday procedural domains such as finding a room, preparing an object or following an administrative sequence. The materials will avoid domain expertise and medical advice, and each item will have an independently reviewed answer key. Two recordings of each instruction will provide acoustic variation without introducing uncontrolled differences in content. Listeners may replay the entire recording or a selected clause; replay controls will be available in every condition so that the adaptive system does not gain an unfair advantage merely by offering basic audio access.

## 3. Planned Evaluation

The pilot will begin with approximately 24 adult volunteers, deliberately selected to span a range of self-reported reading and listening preferences. Short spoken task instructions will be recorded by two speakers and distributed in matched sets. A fixed-format condition and a behavior-adaptive condition will be compared within participant, with items and presentation order counterbalanced.

Accuracy on action-order questions will be the primary endpoint. Replay count, completion time, requests to remove scaffolds and perceived effort (for example the effort item of the NASA-TLX [28]) are secondary. The adaptation policy will be calibrated on a short training set and evaluated on unseen instructions; mixed-effects analysis with crossed random effects for participants and items, and with random slopes justified by the design [29], will account for repeated items. No improvement is assumed, and estimates from a feasibility sample will be reported with intervals rather than treated as definitive efficacy evidence. The caution is warranted by the sample size: pilot estimates of variability are known to be imprecise [30], and simulation work on parallel-group pilots recommends well over the 24 participants planned here for stable variance estimates [31]. The within-participant comparison removes between-listener variance from the contrast, which helps, but the study remains a feasibility pilot.

We will quantify how often the system selects a support level that the listener subsequently rejects, alongside the usual comprehension score. This is needed to distinguish a true gain from a policy that provides assistance so frequently that the participant feels obliged to dismiss it.

Before participant testing, an expert panel will rate pictogram appropriateness and whether the segmented layout preserves order and scope; the pictogram ratings resemble the translucency judgments used in pictogram norming [7]. Disagreement will be resolved through review rather than by choosing the majority image automatically, and agreement among panelists will be summarized with a chance-corrected coefficient such as Krippendorff's alpha [32] so that low-agreement items are flagged for review. For the behavioral experiment, four balanced item lists will rotate across presentation conditions. The same participant will never answer the same question twice. This design sacrifices some within-item repetition in order to avoid an obvious learning effect from repeated exposure to the correct actions.

We will examine task success at the level of both individual steps and complete sequences. A participant who identifies all objects but reverses the action order will not be counted as having understood the instruction. Response time will be measured from the end of the spoken input and interpreted together with replay count. We will also estimate how often the adaptive rule changes the selected format and whether such changes follow a genuine drop in performance. The comparison with fixed support will use the same stimuli, question types and opportunity to request help.

# 4. Discussion

The project should clarify whether adaptive text and visual presentation is useful after spoken input, and which additional costs appear when visual materials are introduced. Clinical populations would require separate recruitment and accessibility review. In particular, variation across an adult convenience sample cannot establish benefit for children with developmental conditions or people with acquired aphasia. Findings will likewise be limited to short, everyday procedural instructions from two speakers; longer discourse, specialized vocabulary or noisy listening conditions may behave differently, and the effect of recognition errors, which would appear on screen as if they had been said, is not examined here.

The principal uncertainty is whether visual support benefits auditory comprehension or simply provides a substitute source of information. Both may be legitimate assistive functions, but they answer different scientific questions. We will report whether participants used audio, text or images when resolving each task, and avoid claims about improved speech perception when the data show only better access to a written representation. Which source resolves each item also matters for the cognitive-load argument: in the dual-channel account of multimedia learning, pictures alongside narration engage separate channels, whereas printed text duplicating narration is the case that account warns against.